\documentclass[conference]{IEEEtran}
\IEEEoverridecommandlockouts

\usepackage[hyphens]{url}
\usepackage{hyperref}
\usepackage[hyphenbreaks]{breakurl}

\usepackage{booktabs}
\usepackage{multirow}
\usepackage{amsmath,amssymb,amsfonts}
\usepackage{graphicx}
\usepackage{amsthm}
\usepackage{url}
\usepackage{listings}
\usepackage{mathtools}
\usepackage[matrix,arrow,curve]{xy}
\usepackage{enumitem}
\usepackage{extarrows}
\usepackage{comment}
\usepackage{tikzsymbols}
\usepackage[utf8]{inputenc}
\usepackage{csquotes}
\usepackage{pbalance}
\usepackage{nowidow}

\newcommand*{\CAMERAREADY}{}%

\ifdefined\COLORDIFF
  \newcommand{\diff}[1]{{\color{cbred} #1}}
\else
  \newcommand{\diff}[1]{{#1}}
\fi

\usepackage{tikz}
\usetikzlibrary{positioning,shapes,arrows,calc,fit,arrows.meta}

\tikzset{
  invisible/.style={opacity=0},
  visible on/.style={alt={#1{}{invisible}}},
  alt/.code args={<#1>#2#3}{%
    \alt<#1>{\pgfkeysalso{#2}}{\pgfkeysalso{#3}} %
  },
}

\usepackage[T1]{fontenc}
\usepackage{amsmath,amssymb}
\usepackage{url}

\newtheorem{example}{Example}

\theoremstyle{definition}
\newtheorem{definition}{Definition}

\theoremstyle{proposition}

\usepackage[numbers,compress]{natbib}

\usepackage[capitalise]{cleveref}

\usepackage{strandstyle}

\newcommand{\myparagraph}[1]{\smallskip\subsubsection*{#1}}
\usepackage{zi4}            %
\usepackage{xcolor}
\usepackage{listings}
\usepackage{lstautogobble}  %
\usepackage{xspace}
\definecolor{dkgreen}{rgb}{0,0.6,0}
\definecolor{ltblue}{rgb}{0,0.4,0.4}
\definecolor{dkviolet}{rgb}{0.3,0,0.5}

\lstdefinelanguage{Coq}{
    mathescape=false,
    texcl=false,
    morekeywords=[1]{Section, Module, End, Require, Import, Export,
        Variable, Variables, Parameter, Parameters, Axiom, Hypothesis,
        Hypotheses, Notation, Local, Tactic, Reserved, Scope, Open, Close,
        Bind, Delimit, Definition, Let, Ltac, Fixpoint, CoFixpoint, Add,
        Morphism, Relation, Implicit, Arguments, Unset, Contextual,
        Strict, Prenex, Implicits, Inductive, CoInductive, Record,
        Structure, Canonical, Coercion, Context, Class, Global, Instance,
        Program, Infix, Theorem, Lemma, Corollary, Coro, llary, Proposition, Fact,
        Remark, Example, Proof, Goal, Save, Qed, Defined, Hint, Resolve,
        Rewrite, View, Search, Show, Print, Printing, All, Eval, Check,
        Projections, inside, outside, Def},
    morekeywords=[2]{forall, exists, exists2, fun, fix, cofix, struct,
        match, with, end, as, in, return, let, if, is, then, else, for, of,
        nosimpl, when},
    morekeywords=[3]{Type,Prop, Set, true, false, option},
    morekeywords=[4]{pose, set, move, case, elim, apply, clear, hnf,
        intro, intros, generalize, rename, pattern, after, destruct,
        induction, using, refine, inversion, injection, rewrite, congr,
        unlock, compute, ring, field, fourier, replace, fold, unfold,
        change, cutrewrite, simpl, have, suff, wlog, suffices, without,
        loss, nat_norm, assert, cut, trivial, revert, bool_congr, nat_congr,
        symmetry, transitivity, auto, split, left, right, autorewrite},
    morekeywords=[5]{by, done, exact, reflexivity, tauto, romega, omega,
        assumption, solve, contradiction, discriminate},
    morekeywords=[6]{do, last, first, try, idtac, repeat},
    morecomment=[s]{(*}{*)},
    showstringspaces=false,
    morestring=[b]",
    morestring=[d]’,
    tabsize=2,
    extendedchars=true,
    inputencoding=utf8,
    sensitive=true,
    breaklines=true,
    basicstyle=\small,
    captionpos=b,
    columns=[l]flexible,
    identifierstyle={\ttfamily\color{black}},
    keywordstyle=[1]{\ttfamily\color{dkviolet}},
    keywordstyle=[2]{\ttfamily\color{dkgreen}},
    keywordstyle=[3]{\ttfamily\color{ltblue}},
    keywordstyle=[5]{\ttfamily\color{dkred}},
    stringstyle=\ttfamily,
    commentstyle={\ttfamily\color{dkgreen}},
    keepspaces,
    xleftmargin=2mm,
    literate=
    {≺}{{$\prec$}}1
    {Σ}{{$\Sigma$}}1
    {ℓ}{{$\ell$}}1
    {Π}{{$\Pi$}}1
    {π}{{$\pi$}}1
    {⊖}{{$\ominus$}}1
    {⊕}{{$\oplus$}}1
    {𝔸}{{$\mathbb{A}$}}1
    {⟨}{{$\langle$}}1
    {⟩}{{$\rangle$}}1
    {⋅}{{$\cdot$}}1
    {ϕ}{{$\phi$}}1
    {ℜ}{{$\mathcal{R}$}}1
    {⊢}{{$\vdash$}}1
    {∈}{{$\in$}}1
    {⊏}{{$\sqsubset$}}1
    {τ}{{$\tau$}}1
    {'}{{$^\prime$}}1
    {forall}{{$\forall$}}1
    {exists}{{$\exists$}}1
    {<-}{{$\leftarrow$}}1
    {=>+}{{$\Rightarrow^+$}}1
    {==}{{\code{==}}}1
    {->}{{$\rightarrow$}}1
    {<->}{{$\leftrightarrow$}}1
    {\#}{{\texttt{\#}}}1
    {\/\\}{{$\wedge$}}1
    {\\\/}{{$\vee$}}1
    {<>}{{$\neq$}}1
    {~}{{$\lnot$}}1
}[keywords,comments,strings]
\newcommand{\arxivonly}[1]{{}}

  \makeatletter
  \def\ps@IEEEtitlepagestyle{
    \def\@oddfoot{\mycopyrightnotice}
    \def\@evenfoot{}
  }

  \def\mycopyrightnotice{
    {\footnotesize
    \begin{minipage}{\textwidth}
    To appear at IEEE CSF'25, June 16-20, 2025, Santa Cruz, CA, USA.
    \copyright~2025 IEEE.
    Personal use of this material is permitted.
    Permission from IEEE must be obtained for all other uses, in any current or future media, including reprinting/republishing this material for advertising or promotional purposes, creating new collective works, for resale or redistribution to servers or lists, or reuse of any copyrighted component of this work in other works.
    The definitive Version of Record is going to appear in the proceedings of the
    38th IEEE Computer Security Foundations Symposium (IEEE CSF'25), June 16-20, 2025, Santa Cruz, CA, USA.
    \end{minipage}
    }
  }

\begin{document}

\newcommand{\easystrands}{\texttt{StrandsRocq}}

\title{
  Strands Rocq:\\  %
  \huge Why is a Security Protocol Correct, Mechanically?
}

\ifdefined\CAMERAREADY
  \author{
      \IEEEauthorblockN{Matteo Busi}
      \IEEEauthorblockA{
      \textit{DAIS, Ca' Foscari University}\\
      Venice, Italy \\
      matteo.busi@unive.it}
  \and
      \IEEEauthorblockN{Riccardo Focardi}
      \IEEEauthorblockA{
      \textit{DAIS, Ca' Foscari University}\\
      Venice, Italy \\
      focardi@unive.it}
  \and
      \IEEEauthorblockN{Flaminia L. Luccio}
      \IEEEauthorblockA{
      \textit{DAIS, Ca' Foscari University}\\
      Venice, Italy \\
      luccio@unive.it}
  }
\else
  \author{Anonymous author(s)}
\fi

\maketitle

\newcommand{\enc}[2]{{\ensuremath {\langle #1 \rangle _{#2}}}}

\begin{abstract}
Strand spaces are a formal framework for symbolic protocol verification that allows for pen-and-paper proofs of security \cite{FHG98}. While extremely insightful, pen-and-paper proofs are error-prone, and it is hard to gain confidence on their correctness. To overcome this problem, we developed \easystrands, a full mechanization of the strand spaces in Coq (soon to be renamed Rocq). The mechanization was designed to be faithful to the original pen-and-paper development, and it was engineered to be modular and extensible. \easystrands{} incorporates new original proof techniques, a novel notion of maximal penetrator that enables protocol compositionality, and a set of Coq tactics tailored to the domain, facilitating proof automation and reuse, and simplifying the work of protocol analysts. To demonstrate the versatility of our approach, we modelled and analyzed a family of authentication protocols, drawing inspiration from ISO/IEC 9798-2 two-pass authentication, the classical Needham-Schroeder-Lowe protocol, as well as a recently-proposed static analysis for a key management API. The analyses in \easystrands{} confirmed the high degree of proof reuse, and enabled us to distill the minimal requirements for protocol security. Through mechanization, we identified and addressed several issues in the original proofs and we were able to significantly improve the precision of the static analysis for the key management API. Moreover, we were able to leverage the novel notion of maximal penetrator to provide a compositional proof of security for two simple authentication protocols.
\end{abstract}


\section{Introduction}

The literature on the analysis of cryptographic protocols is extensive and highly diverse, as evidenced by comprehensive surveys such as~\cite{barbosa:SoKCAC,blanchetPOST2012,CortierSurvey2011}.
Strand spaces have been a pioneering formalism for the specification and analysis of security protocols~\cite{FHG98}.
The slogan in the title, ``why is a security protocol correct?'' that we borrowed and extended in our paper, succinctly captures the underlying motivation: strand spaces were crafted to enable intuitive reasoning about protocol security.
Impressively, they facilitate concise and insightful pen-and-paper proofs of security for protocols featuring unbounded participants, sessions, keys, nonces, and more.
In the process of proving security, it becomes natural to introduce an assumption only when needed, resulting in a set of minimal assumptions necessary for the security proof to hold.
This approach is extremely insightful as it guides the analyst to gain a deep understanding of the root reasons behind protocol security and the crucial assumptions for such security to hold.
To the best of our knowledge, no other formalism allows for this level of insightfulness.
In fact, strand spaces had a significant impact on the research community, leading to a considerable amount of follow-up work and extensions.
Notable examples include~\cite{CVB05,CDLMS03,HP03,KL09,YEMMS16}, just to mention a few.

In the era of automated and mechanized verification, interest in pen-and-paper proofs is waning.
While strand spaces allow for concise proofs, the manual analysis of complex protocols is not credible, and even experimenting with variants of the same protocol can become tedious and time-consuming.
Moreover, mistakes can occur in pen-and-paper proofs.
The Cryptographic Protocol Shapes Analyzer (CPSA) \cite{cpsa,LRGR16} offers automated verification of protocols based on strand spaces.
However, it lacks the beauty and insightfulness of pen-and-paper proofs, and has been overshadowed by mainstream popular tools (see, e.g., \cite{BSCS20,EMM09,MSCB13}).
The primary motivation of this work is to revitalize the strand space model.
While our focus remains on understanding ``why is a security protocol correct?'' we aim to achieve this in a mechanized and reusable manner, by providing extensions that lessen the effort required by analysts to write security proofs.
In particular, we aim for: full mechanization; a good degree of proof automation to keep proofs short, readable, and reusable;
compositionality results to improve tool scalability.

In this paper, we present \easystrands{},\footnote{pronounced ``\emph{Strands Rock!}''.} the first full implementation of the strand spaces model in the Coq proof assistant.
Coq is extremely appealing for our goal as it offers full flexibility, allows for fully mechanized proofs with a small and popular trusted computing base, and provides the possibility of developing tactics for proof automation.
The development of the precise strand spaces model in Coq was challenging and required the development of original proof techniques and tactics to achieve a satisfactory degree of proof automation, eliminating all trivial and tedious cases.
A significant effort was also put in engineering the library to ensure its reusability for different protocols.

As confirmation, we mechanized several security proofs.
We began with a family of simple authentication protocols inspired by the ISO/IEC 9798-2 two-pass authentication protocol~\cite{ISO97982}, which we successfully analyzed in five different variants while significantly reusing the proofs. This confirmed that, although the initial effort to analyze a new protocol may be greater than with popular automated tools, once a proof is established, \easystrands{} allows for exploring protocol variants with relatively low effort. Furthermore, we introduced the notion of \emph{maximal penetrator} based on restrictions regarding sensitive cryptographic operations rather than enumerating all possible malicious capabilities. This approach enables the composition of two protocols proven secure under their respective maximal penetrators if they adhere to each other's conditions. We applied this technique to two simple authentication protocols, successfully proving the security of their composition by reusing the individual proofs.

We reproduced the pen-and-paper proofs of the classic Needham-Schroeder-Protocol (NSL) from the original strand spaces paper \cite{FHG98}, using exactly the same arguments and proof techniques, and then improved and simplified them using the new proof techniques offered by \easystrands{}.
We also verified the results of \citet{focardi2021secure}, correcting some errors in the original pen-and-paper development and improving their findings. Through mechanization, we identified some superfluous conditions in the proposed static analysis, which we refined to enhance its precision. To validate our approach, we demonstrate that the most complex example presented by \citet{focardi2021secure}, originally used to illustrate the limitations of their analysis, can be proven correct using our refined solution.

\myparagraph{Main contributions}
We summarize our main contributions as follows:
\begin{itemize}
\item we provide the first fully mechanized implementation of the original strand spaces model in a proof assistant (\cref{sec:running}) and we implement Coq libraries that allow for the automation of case analysis in the proofs (\cref{sec:proofprotocol}), making them concise and reusable (\cref{sec:reusing});
\item we devise new proof techniques that overcome some limitations of the ones used in \cite{FHG98} (\cref{sec:newproofs}) and provide protocol compositionality through the notion of maximal penetrator (\cref{sec:maximal});
\item we analyze a family of simple authentication protocols inspired by the ISO/IEC 9798-2 two-pass authentication protocol~\cite{ISO97982}, across several different variants, with significant proof reuse,  providing insights into the minimal security assumptions required for each variant.
\item we reproduce the analysis of the NSL protocol from~\cite{FHG98}, addressing a few mistakes in the pen-and-paper lemmas and proofs (\cref{subsec:NSL});
\item we mechanize and fix a recent proof of security for a key management API based on a static analysis of the policy~\cite{focardi2021secure}, a
task we believe to be particularly unsuitable for automated tools, since the security theorem is based on an overapproximation of the API behavior (\cref{subsec:KMP}). We propose an enhanced static analysis, demonstrating its security while largely reusing our mechanized proof for the original analysis.
\end{itemize}

\ifdefined\CAMERAREADY
  \myparagraph{Note} The Coq implementation, the examples and case studies are available online~\cite{strandsrocqcode}. 
\else
\myparagraph{Note} The Coq implementation, the examples and case studies are fully available to the reviewers as supplementary material.
\fi

\section{Related Work}

\myparagraph{Mechanized protocol analysis}
The literature on mechanized protocol analysis from the past two decades is extensive and challenging to encompass within a single paragraph (see, e.g., \cite{barbosa:SoKCAC,blanchetPOST2012,CortierSurvey2011}). Researchers have studied various methodologies to prove the correctness of security protocols, both in the idealized world of symbolic models and in the more concrete realm of computational models. These two approaches complement each other and are sometimes connected by computational soundness results that allow for obtaining proofs in computational models through purely symbolic analysis \cite{blanchetPOST2012}.
Symbolic analysis has scaled to the point of automatically verifying real-world protocols, and the plethora of popular tools, such as \cite{cpsa,BSCS20,EMM09,MSCB13,Paulson98}, confirms the success
of this approach.

Other successful approaches in the literature offer semi-automated, interactive symbolic techniques in which proofs are partially provided by hand. A representative example is DY* \cite{DY}, a recently proposed verification framework for symbolic protocol analysis based on F* \cite{Fstar}. DY* leverages dependent types to prove protocol security, and allows for extracting protocol implementations in F*. Interestingly, Bhargavan et al. \cite{DY} point out how automated tools impose limitations on the protocol model to keep the analysis feasible, and how some protocols require powerful inductive reasoning provided by general-purpose proof frameworks such as Coq \cite{coq} and F* \cite{Fstar}, which are only partially supported by state-of-the-art automated tools. In this work, following the direction set by DY*, aiming to provide full flexibility of general-purpose proof frameworks for symbolic protocol analysis,  we fully mechanize strand spaces in Coq, and provide mechanized and reusable proofs of security for significant core examples. We are certainly far from competing with state-of-the-art tools in terms of coverage and scalability, but we claim that our contribution is significant and distinguished from various perspectives that we discuss below.

\myparagraph{Mechanized proof methods for symbolic protocol analysis}
The model proposed by Paulson \cite{Paulson98}, based on the Isabelle theorem prover \cite{Paulson94}, is the closest to our mechanization of strand spaces. Interestingly, both approaches rely on inductive reasoning, and Fabrega et al. \cite{FHG98} briefly discuss the differences between them pointing out that strand spaces have a peculiar underlying causal semantics that represents protocol executions as partially ordered events. This allows for a powerful inductive principle that permits to prove properties of executions by reasoning on the local behaviour of protocol participants. Moreover, this provides very insightful proofs of protocol security, which point out necessary conditions, such as nonce freshness and key secrecy, in a direct and very intuitive way, allowing to prove protocol security under a set of minimal necessary conditions.
Notably, Paulson's approach~\cite{Paulson98}  has been applied to complex and realistic protocols, demonstrating its scalability through proof automation in Isabelle (see, e.g., \cite{BellaP98}). The lack of automation for strand spaces proofs has historically hindered achieving similar results and conducting a close comparison between the two approaches. We anticipate that \easystrands{} will address this gap and offer further insights into their relative merits, similarities, and differences.

Another closely related work is DY* \cite{DY} that we already mentioned above.
While our proposal is certainly not as scalable or mature as DY*, we believe it offers interesting distinctions. Specifically, in \easystrands, we demonstrate protocol security on straightforward specifications that directly correspond to the so-called \emph{Alice and Bob notation}.
In fact, our
security proofs are conducted on the plain protocol specification without any typing annotations.
While in DY*, the primary analyst effort is on providing sophisticated dependent types to enable protocol typechecking, our approach amounts to devising a suitable property that is proven inductively.
For this reason, we believe that both approaches offer valuable insights and complement each other. However, the maturity of DY* and F* is so significantly higher compared to our approach that it is nearly impossible to make a direct comparison between the two.

\myparagraph{Mechanization of strand spaces}
Li and Pang \cite{li2013inductive} proposed a version of strand spaces in Isabelle/HOL.
The authors partially modified and extended the original strand spaces model \cite{FHG98}, basing their inductive verification on the concept of authentication tests from \cite{guttman2000authentication}.
They highlighted that their automated proof of the Needham-Schroeder-Lowe (NSL) protocol spans 1954 lines, and the proof scripts are still available at \cite{Yongjianstrands}.
In contrast, our goal was to provide an accurate mechanization of the original strand spaces model without resorting to any modification or extension. Despite the initial intricacy of the process, we successfully mechanized strand spaces with no alterations. As confirmation of our successful mechanization, we were able to reproduce the exact pen-and-paper proofs for the NSL protocol (\cref{subsec:NSL}), which, in our case, are much shorter than those presented in \cite{li2013inductive}, totaling nearly 400 lines for the verification of the same properties. Additionally, we developed Coq tactics that enable proof automation and reuse, as discussed in \cref{sec:proofprotocol,sec:reusing}, and
devised new proof techniques that overcome some limitations of the ones used in \cite{FHG98} (\cref{sec:newproofs}), providing protocol compositionality through the notion of maximal penetrator (\cref{sec:maximal}).

The Cryptographic Protocol Shapes Analyzer (CPSA) is a tool designed for the analysis and design of security protocols grounded in strand space theory~\cite{cpsa,LRGR16}, while Maude-NPA adopts the strand spaces formalism to formally specify protocols~\cite{EMM09}.
Both tools have been employed for the automated analysis of various cryptographic protocols and security APIs \cite{MaudeAPI1,MaudeAPI2,J12,LZ17,RGMO12,RDGR18,SRP20}, and they belong to the category of automated symbolic verification tools. Our focus is complementary, as we do not aim for fully automated verification.

Strand spaces rely on nonstandard definitions and introduce a specific induction principle, necessitating careful formalization and tailored proof automation techniques to prevent proofs from becoming lengthy and tedious, thus avoiding distractions for the analyst with too many uninteresting cases. This presented a significant challenge, and our mechanization of strand spaces in Coq is the first comprehensive one, including significant examples.
We found existing repositories lacking protocol examples and seemingly unmaintained \cite{kentstrands,nguyenauthtests,nguyenstrands}. The only related published document is a BSc thesis by Hai Hoang Nguyen \cite{nguyenthesis}, which is related to the aforementioned repositories.

\section{Background on Strand Spaces}\label{sec:background-new}
  In this section, we review the primary components of strand spaces from \cite{FHG98}.

\myparagraph{Strands and terms} Intuitively, a \emph{strand} is an ordered sequence of events, denoting the activity of either a legitimate participant in a security protocol or a series of actions performed by an intruder.
 The events consist of a term $t \in \terms$ being
 transmitted $+t$  or received $-t$.
 The set of all possible signed terms is denoted $\pm \terms$, and a finite sequence of such events is called a {\em strand}, denoted $s$, and is an element of $(\pm \terms)^*$.
 A collection of strands is called a \emph{strand space}, and it includes strands of the legitimate participants and strands of the penetrator.

We let $\T$ be a set of atomic messages (texts), and $\termKey$ be a set of cryptographic keys, disjoint from $\T$, equipped with a function $\inv: \termKey \rightarrow \termKey$, providing the inverse of a given key. The function $\inv$ associates each element of a key pair in an asymmetric cryptosystem with its counterpart, and it associates a symmetric key with itself. We write $k^{-1}$ to denote $\inv(k)$. Then, we write $\enc{g}{k}$ to denote term $g$ encrypted under key $k$ and  $g \cdot h$ to denote the concatenation of terms $g$ and $h$.
Finally, we let \terms denote the set of all terms constructed by applying encryption and concatenation starting from $\T$ and $\termKey$. The subterm relation $\sqsubset$ is used to express that a certain term occurs into another one, and thus in the corresponding node.
Notice that, the subterm relation does not consider a cryptographic key $k$ in a ciphertext \enc{.}{k} as a subterm of the ciphertext, given that $k$ does not occur in the message payload but it is instead used to generate the ciphertext.

 \begin{example}[Simple authentication protocol]
  \label{ex:simpleprotocol}
  We consider a simple unilateral authentication protocol based on symmetric key cryptography and nonces, inspired from
  ISO/IEC 9798-2 two-pass authentication~\cite{ISO97982}:
  \begin{align*}
    A \rightarrow B & : A \cdot B \cdot N_a \\
    B \rightarrow A & : \enc{N_a \cdot A}{ \SK{AB}}
  \end{align*}
Intuitively, the protocol begins with Alice ($A$) initiating communication by sending the principal identifiers and a fresh nonce $N_a$ to Bob ($B$). Bob encrypts the nonce $N_a$ along with the identifier of $A$ using a  symmetric key $\SK{AB}$ shared between $A$ and $B$. Alice then verifies that the received message is indeed encrypted under the correct key and includes the nonce $N_a$ along with her identifier $A$. This confirmation is sufficient to convince her that she is communicating with Bob, a security guarantee referred to as \emph{unilateral authentication}.
Notice that, $A$ and $B$ in the initial message constitute Alice's initial claim regarding the protocol session. However, they lack security significance as they can be manipulated by the attacker. Consequently, the security of the protocol does not rely on them.

The protocol is formalized in the strand spaces model by specifying that
the initiator strands have the form
\[ \strand{ + A \cdot B \cdot N_a; - \enc{N_a \cdot A}{ \SK{AB}} } \]
while the responder strands have the form
\[ \strand{ - A \cdot B \cdot N_a; + \enc{N_a \cdot A}{ \SK{AB}} } \]
for all $A, B, N_a \in \T$.
\end{example}
\myparagraph{Nodes, bundles, and the penetrator}
A strand space has an associated graph in which nodes are assigned to events in a strand and are indexed by the event position. A node is, in fact, noted as a pair $\node{ s, i }$ representing the node association to the $i$-th event of the strand name $s$. The graph includes edges connecting output and input events related to the same message, as well as consecutive events on the same strand. In particular, every node $\node{ s, i }$ with an output event $+t$ is related to node $\node{ s', i' }$ with the corresponding input event $-t$ through the \emph{interstrand} relation $\node{ s, i } \rightarrow \node{ s', i' }$. Moreover, each node $\node{ s, i }$ is related to the next one $\node{ s, i+1 }$ in the same strand through the \emph{intrastrand} relation $\node{ s, i } \Rightarrow \node{ s, i+1 }$. We use $\term(n)$ to denote the event (i.e., the signed term) associated with node $n$.

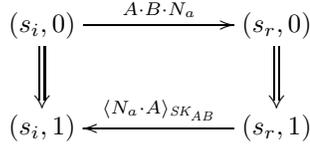
\begin{figure}[t]
  $$\xymatrix@R=3mm@C=20mm{
  \node{s_i,0}\ar@{=>}[dd]\ar[r]^{A \cdot B \cdot N_a}&\node{s_r,0}\ar@{=>}[dd]\\ \\
  \node{s_i,1}&\node{s_r,1}\ar[l]_{\enc{N_a \cdot A}{ \SK{AB}}}\\
  }$$
  \caption{A bundle for the unilateral authentication protocol of \cref{ex:simpleprotocol}.}
  \label{fig:simpleprotocol}
\end{figure}

 Protocol runs are modeled by \emph{bundles}, which select events from the strand space and display their causal dependencies, establishing a partial order for events in the run.
 Bundles are finite and acyclic subgraphs of the strand space graph. Each event within a bundle requires all preceding events on the same strand, along with the corresponding edges indicating strand precedence. Every input event in the bundle is linked by a single incoming edge from an output event.

\begin{example}[Bundle]
  \label{ex:bundle}
Consider again the protocol of \cref{ex:simpleprotocol}.
A bundle for a given instance of $A, B$ and $N_a$ is depicted in \cref{fig:simpleprotocol}.
On the left, we see the two nodes \node{s_i,0}, \node{s_i,1} of the initiator strand $s_i$, and on the right, the two nodes \node{s_r,0}, \node{s_r,1} of the responder strand $s_r$ connected vertically by the intrastrand relation $\Rightarrow$. This represents the causal dependencies between the sequential events in each strand. Then, we see the interstrand relation $\rightarrow$ connecting outputs and inputs: the initiator sends the message $A \cdot B \cdot N_a$ to the responder, who answers with $\enc{N_a \cdot A}{ \SK{AB}}$. Notice, in particular, that the events, i.e., the signed terms, associated to the nodes are:
\[
  \begin{array}{lll}
\term(~\node{s_i,0}~) & = & + A \cdot B \cdot N_a\\
\term(~\node{s_i,1}~) & = & - \enc{N_a \cdot A}{ \SK{AB}} \\
\term(~\node{s_r,0}~) & = & - A \cdot B \cdot N_a\\
\term(~\node{s_r,1}~) & = & + \enc{N_a \cdot A}{ \SK{AB}}
  \end{array}
  \]
\end{example}
\smallskip
Penetrator strands model a standard Dolev-Yao attacker that  intercepts, duplicates, and manipulates messages, knows a subset $\KP$ of the keys $\termKey$, and encrypts and decrypts messages only when they know the appropriate key.
\begin{definition}[Penetrator \cite{FHG98}]
\label{def:penetrator}
Let $g, h, m \in \terms$ note generic terms, and $k \in \termKey$ a key.
A penetrator strand has one of the following forms:
\begin{description}[leftmargin=9.5em,style=nextline]
\item[~~\rm Text message] \strand{+t} with $t \in \T$
\item[~~\rm Flushing] \strand{-g}
\item[~~\rm Tee] \strand{-g;~ +g ;~ +g}
\item[~~\rm Concatenation] \strand{-g;~ {-h};~ {+g \cdot h}}
\item[~~\rm Separation] \strand{-g \cdot h;~ {+g};~ {+h}}
\item[~~\rm Key] \strand{+k} with $k \in \KP$
\item[~~\rm Encryption] \strand{-k;~ {-m};~ {+\enc{m}{k}}}
\item[~~\rm Decryption] \strand{-k^{-1};~ {-\enc{m}{k}};~ {+m}}
\end{description}
\end{definition}

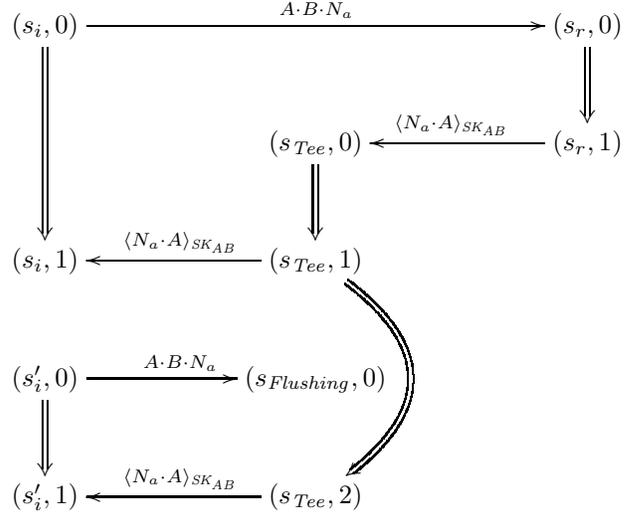
\begin{figure}[t]
  $$\xymatrix@R=10mm@C=20mm{
  \node{s_i,0}\ar@{=>}[dd]\ar[rr]^{A \cdot B \cdot N_a} & & \node{s_r,0}\ar@{=>}[d]\\
   & \node{s_{\mathit{Tee}},0}\ar@{=>}[d] & \node{s_r,1}\ar[l]_{\enc{N_a \cdot A}{ \SK{AB}}}\\
  \node{s_i,1}  & \node{s_{\mathit{Tee}},1}\ar[l]_{\enc{N_a \cdot A}{ \SK{AB}}}\ar@{=>}@/^3pc/[dd] \\
  \node{s'_i,0}\ar@{=>}[d]\ar[r]^{A \cdot B \cdot N_a} & \node{s_{\mathit{Flushing}},0}\\
  \node{s'_i,1} & \node{s_{\mathit{Tee}},2}\ar[l]_{\enc{N_a \cdot A}{ \SK{AB}}}\\
  }$$
  \caption{A bundle representing a replay attack when there is no assumption on $N_a$ freshness. The attack is prevented by requiring that $N_a$ uniquely originates in \node{s_i,0}.}
  \label{fig:replay}
\end{figure}

\begin{example}[A replay attack]
  For the protocol of \cref{ex:simpleprotocol} we assume that $\SK{AB} \not\in \KP$ in order to prevent trivial attacks in which the penetrator knows $\SK{AB}$. However, this assumption is not enough: the protocol prevents replay attacks thanks to the freshness of the nonce $N_a$. We have not formalized this assumption yet, which means that the attack is possible, as illustrated in \cref{fig:replay}. There are two initiator strands $s_i$, $s'_i$ using the same nonce $N_a$. The penetrator intercepts the message $\enc{N_a \cdot A}{ \SK{AB}}$ from the responder and replays it twice thanks to the \emph{Tee} strand $s_{\mathit{Tee}}$. Before sending the second copy, the penetrator drops the first message $A \cdot B \cdot N_a$ from the second initiator $s'_i$ using a \emph{Flushing} strand $s_{\mathit{Flush}}$.

  Nonce freshness is formalized by requiring that $N_a$ \emph{uniquely originates} in node \node{s_i,0}. Intuitively, this means that if $N_a$ appears in a positive node of a strand and there is no preceding node in the same strand containing $N_a$, then such a node must be \node{s_i,0}. This condition makes the bundle of \cref{fig:replay} invalid since $N_a$ originates both on \node{s_i,0} and on \node{s'_i,0}. In fact, the freshness of $N_a$ is a necessary condition to prevent replay attacks on this particular protocol.
\end{example}

\myparagraph{Proving security}
Strand spaces can be used to formalize and prove various security properties. For example, unilateral authentication can be expressed as a standard
agreement statement~\cite{lowe1997hierarchy}: for any initiator strand $s_i$ with parameters $A, B, N_a$ in a given bundle, there exists a responder strand $s_r$ that agrees on $A, B, N_a$.
The proof technique introduced in \cite{FHG98} is elegant and effective and leverage the fact that given a bundle $C$ the reflexive and transitive closure of the two relations $\rightarrow$ and $\Rightarrow$ that are part of $C$ define a partial order $\preceq_C$. From this, it is possible to prove that any nonempty subset of $C$ has a $\preceq_C$-minimal element. This provides a powerful inductive principle that can be applied locally on each protocol and penetrator strand.

\begin{example}[Proof sketch for agreement]
  \label{ex:proofs}
  In order to prove agreement for the protocol of \cref{ex:simpleprotocol}, i.e., that a responder strand $s_r$ exists and it agrees on $A, B$ and $N_a$, we consider the set $S = \{ m \in  C : \enc{N_a \cdot A}{ \SK{AB}} \sqsubset term(m) \}$ for a given bundle $C$. We know that this set is nonempty because $term(~\node{s_i,1}~) = \enc{N_a \cdot A}{ \SK{AB}}$ and so $\node{s_i,1} \in S$. So, as discussed before, $S$ has a $\preceq_C$-minimal element.
  The intriguing observation now is that it is possible to prove that the minimal element must be \node{s_r,1} (see \cref{fig:simpleprotocol}), i.e., the last node of the responder strand $s_r$, thereby establishing the agreement result, by a purely local reasoning over strands.
  It is enough to examine each individual strand to demonstrate that  \enc{N_a \cdot A}{ \SK{AB}} cannot \emph{originate}, i.e., appear for the first time, in any of them, except precisely in \node{s_r,1}.

  For example, consider the {Concatenation} strand
  $ \strand{-g;~ {-h};~ {+g \cdot h}} $ of the penetrator. Minimal elements are always positive due to the fact that negative nodes in a bundle are always preceded, via $\rightarrow$ by a positive node with the same term.
  The only positive node has term ${g \cdot h}$. Suppose this is minimal in $S$. Then, $\enc{N_a \cdot A}{ \SK{AB}} \sqsubset {g \cdot h}$ which implies that either $\enc{N_a \cdot A}{ \SK{AB}} \sqsubset g$ or $\enc{N_a \cdot A}{ \SK{AB}} \sqsubset h$. In both cases we get a contradiction as respectively the first or the second node of the strand would belong to $S$, breaking the minimality assumption. A fully mechanized proof for this protocol will be presented in \cref{sec:proofprotocol}.
\end{example}

\section{Mechanizing Strand Spaces: \easystrands}\label{sec:running}
  
In this section,
we present \easystrands, a complete mechanization of strand spaces in Coq.
We briefly introduce the structure and engineering of the library (\cref{sec:library}). Then, we demonstrate the process of specifying and proving the correctness of the protocol presented in \cref{ex:simpleprotocol} through simple steps, illustrating the specification phase (\cref{sec:specification}), the underlying proof technique, its mechanization, and our novel proof automation techniques (\cref{sec:proofprotocol}) that allow for compact and reusable proofs (\cref{sec:reusing}).
Finally in \cref{sec:newproofs} we present a new proof technique that simplifies the one presented by Fabrega et al. \cite{FHG98}.
During this journey, we start with the basic authentication protocol from \cref{ex:simpleprotocol}, inspired by the ISO/IEC 9798-2 two-pass authentication protocol~\cite{ISO97982}, and successfully analyze five different variants, uncovering the minimal security assumptions for each of them.

\ifdefined\CAMERAREADY
  \myparagraph{Note for the readers} The complete mechanization and proofs are available online~\cite{strandsrocqcode}.
\else
  \myparagraph{Note for the reviewers} The complete mechanization and proofs are included as supplementary material with the submission.
\fi

\subsection{The \easystrands{} library}
\label{sec:library}
We organized the library into modules, separating the theory of strands based on abstract domains, as in the original paper (folder \lstinline|Common|), from an implementation that we believe is more convenient for verifying protocols.
Implementing the abstract domains is an important sanity check to remove all axioms and assumptions, ensuring that such assumptions are realistic (folder \lstinline|Instances|).
For example, concrete terms are part of \lstinline|Instances|, which makes the library very flexible if one wishes to model new cryptographic primitives: the entire \lstinline|Common| section remains unchanged, and it is only necessary to instantiate a specific \lstinline|Module Type|.
Unlike the abstract strand definition from~\cref{sec:background-new}, strands are instantiated here as \lstinline|Σ := nat * list sT|.
The natural number serves as a strand identifier and \lstinline|list sT| is a list of signed terms denoting the trace associated with the strand.
This choice is particularly convenient for protocol specification as it allows for specifying strands and their traces in a single place.
In the original paper, traces are bound to strands through a separate function \lstinline|tr|.
In our implementation we just have that \lstinline|tr s| is defined as \lstinline|snd s|, i.e., the second field of the strand instantiation.

\subsection{Modelling Protocols}
\label{sec:specification}

We define the roles in the protocol by inductively listing all the possible strands they can undertake.
This might seem overly intricate since, in most cases, honest principals follow a single execution trace that is quantified over parameters and payload values.
Nevertheless, in general, a principal could engage in more than one trace.
For instance, a penetrator may carry out various potential traces (\cref{sec:background-new}).
Additionally, when modeling key management APIs (\cref{subsec:KMP}), a single principal/device can implement various functionalities, each represented by a distinct trace.

Starting now, we directly present the notation employed in \easystrands, which deviates slightly from the mathematical notation used so far.
We use \lstinline{Na} to represent the nonce $N_a$, \lstinline{SK A B} to denote the key $\mathit{\SK{AB}}$, \lstinline{⟨ M ⟩_(K)} to indicate $\enc{M}{K}$, and \lstinline{⊕}, \lstinline{⊖} to respectively denote $+$ and $-$.
Since the type of \lstinline{A}, \lstinline{B} and \lstinline{Na} is \T, representing atomic terms, we respectively write \lstinline{$A}, \lstinline{$B} and \lstinline{$Na} to represent their values as general terms of type \terms.
For the protocol of \cref{ex:simpleprotocol},
the initiator strands are defined as follows:
\begin{lstlisting}
Inductive SA_initiator_strand (A B Na : T) : Σ -> Prop :=
  | SAS_Init : forall i,
      SA_initiator_strand A B Na
        (i, [ ⊕ $A ⋅ $B ⋅ $Na; ⊖ ⟨ $Na ⋅ $A ⟩_(SK A B) ]).
\end{lstlisting}
Dually, the responder strands have swapped inputs and outputs:
\begin{lstlisting}
Inductive SA_responder_strand (A B Na : T) : Σ -> Prop :=
  | SAS_Resp : forall i,
      SA_responder_strand A B Na
        (i, [ ⊖ $A ⋅ $B ⋅ $Na; ⊕ ⟨ $Na ⋅ $A ⟩_(SK A B) ]).
\end{lstlisting}
To analyze this protocol we will restrict ourselves to strands of three types: \lstinline{penetrator_strand} (defined along \cref{sec:background-new}), \lstinline{SA_initiator_strand}, or \lstinline{SA_responder_strand}:
\begin{lstlisting}
Inductive SA_StrandSpace (K__P : K -> Prop) : Σ -> Prop :=
  | SASS_Pen  : forall s,
    penetrator_strand K__P s -> SA_StrandSpace K__P s
  | SASS_Init : forall (A B Na : T) s,
    SA_initiator_strand A B Na s -> SA_StrandSpace K__P s
  | SASS_Resp : forall (A B Na : T) s,
    SA_responder_strand A B Na s -> SA_StrandSpace K__P s
\end{lstlisting}
where \lstinline|K__P| encodes the knowledge of the penetrator at the beginning of the execution.
For our purposes the following minimal definition suffices:
\begin{lstlisting}
Definition K__P_AB (A B : T) (k : K) := k <> SK A B.
\end{lstlisting}
Intuitively, we assume that the only key the penetrator should not know is the actual key used by the two honest parties.

\subsection{Proof Automation}
\label{sec:proofprotocol}

We have developed a Coq library and some tactics to efficiently implement case analysis over strands, searching for a minimal element over a given strand.
We illustrate their usage below.
From now on we assume to have two honest parties \lstinline|A| and \lstinline|B|, a nonce \lstinline|Na| and a bundle \lstinline|C| whose nodes belong to the protocol strands \lstinline|SA_StrandSpace (K__P_AB A B)| in which the attacker does not know the key \lstinline|SK A B|. Since we want to prove authentication for the initiator, we assume that \lstinline|C| contains an initiator strand \lstinline|s| with the appropriate parameters, i.e., \lstinline|SA_initiator_strand A B Na s|.
All of these variables and hypotheses are specified locally using Coq \lstinline|Variable| and \lstinline|Hypothesis| commands and make propositions and lemmas more succinct and readable.

We consider \emph{non-injective agreement} requiring that, under the above assumptions,
there exists a responder strand \lstinline{s'}, and the initiator and responder traces agree on parameters
\lstinline{A}, \lstinline{B} and \lstinline{Na}.
Formally,
\begin{lstlisting}
Proposition noninjective_agreement :
  exists s' : Σ,
    SA_responder_strand A B Na s' /\ is_strand_of s' C.
\end{lstlisting}

\noindent
As illustrated in \cref{ex:proofs},
the proof in the strand spaces model revolves around showing that only the responder, with parameters \lstinline{A}, \lstinline{B} and \lstinline{Na}, can generate the expected ciphertext \lstinline{c = (⟨ $Na ⋅ $A ⟩_(SK A B))}.
The proof
is based on lemma \texttt{\small exists\_minimal\_bundle} (see \lstinline{Common/Bundles.v}) stating that each nonempty subset of nodes has a minimal w.r.t.\ the $\preceq_C$ relation (\cref{sec:background-new}).

The proof inspects all possible kinds of strands for \lstinline{s}: penetrator, initiator and responder.
Doing this in Coq is tedious and requires repetitive proofs even for cases that are deemed as trivial in pen-and-paper proofs.
For this reason, \easystrands{} provides a characterization of the minimal element of set of nodes in terms of a logical proposition covering all the possible cases.
For example, for the first strand of the penetrator, which is the output of an atomic term \lstinline{t} written \lstinline{[⊕ $t]} we obtain the
proposition \lstinline{False \/ c = $t /\ True /\ index m = 0}
which is false since \lstinline{c} is a ciphertext and it cannot be that \lstinline{c = $t}.
Other cases are more complex, e.g., for pair generation \lstinline{[⊖ g; ⊖ h; ⊕ g ⋅ h]} we get
\begin{lstlisting}
((False \/ subterm c g /\ False /\ index m = 0) \/
~ subterm c g /\ subterm c h /\ False /\ index m = 1) \/
~ subterm c g /\ ~ subterm c h /\
(c = g ⋅ h \/ subterm c g \/ subterm c h) /\
True /\ index m = 2
\end{lstlisting}
that is less trivial to analyze manually.
Therefore, we have implemented a tactic called \lstinline{simplify_prop}, which recursively simplifies propositions, leveraging the decidability of underlying statements.
It also attempts to automatically prove straightforward
(in)equalities, such as \lstinline{c <> $t} in the first case of the penetrator.

When applied to the penetrator case, the \lstinline{simplify_prop} tactic eliminates seven out of eighth cases, leaving only the interesting one, i.e., the encryption case with trace
\begin{lstlisting}
[⊖ #(SK A B); ⊖ $Na ⋅ $A; ⊕ (⟨ $Na ⋅ $A ⟩_(SK A B))]
\end{lstlisting}
Intuitively, this refers to the case where the penetrator generates the ciphertext \lstinline{c}, which is used by \lstinline{A} to confirm the identity of \lstinline{B}.
We eliminate this case by exploiting the fact that the penetrator can never learn a secure symmetric key.
This can be proved using a general property regarding the penetrator,
which asserts that the key read in the first node of a penetrator's encryption strand, in this case \lstinline{SK A B}, cannot be equal to a key that is not initially known by the penetrator and does not originate on a honest participant strand.
The fact that \lstinline{SK A B} is not initially known by the attacker is a direct consequence of the definition of \lstinline{K__P_AB A B} as \lstinline|k <> SK A B|. Additionally, the fact that \lstinline{SK A B} is not generated by the honest participants is demonstrated through a simple lemma, which can be proved using the same proof automation technique in just 8 lines of Coq.
So, we conclude that it must be \lstinline{SK A B <> SK A B}, leading to a contradiction.

The initiator case is automatically resolved by the \lstinline{simplify_prop} tactic, while the responder case leaves us with two subcases, depending on whether \lstinline{A} and \lstinline{B} are equal or not.
Both cases are resolved easily, as they yield a valid binding for the protocol parameters.
Interestingly, thanks to our proof automation techniques, the whole proof of \lstinline|noninjective_agreement| amounts to about 60 lines, as is greatly reusable as we will se next.

We also prove that each responder session corresponds to a different initiator session, i.e., that authentication is \emph{injective} and cannot be reused in a replay attack.
\begin{lstlisting}
Proposition injectivity :
  uniquely_originates $Na ->
    forall U U' s',
      SA_initiator_traces U U' Na (tr s') -> s' = s.
\end{lstlisting}
Notice that this property only holds if \lstinline{Na} is freshly generated which, in the strand spaces model, is captured by the \lstinline{uniquely_originates} definition stating that \lstinline{Na} originates, i.e., appears for the first time, in a unique node in a given bundle.
Injective agreement follows as a corollary from \lstinline{noninjective_agreement} and \lstinline{injectivity} (see \lstinline{injective_agreement} in \lstinline{Examples/simple_auth/SimpleAuth.v}).
\subsection{Proof Reuse}\label{sec:reusing}
An important feature of protocol analysis tools is the ability to \emph{play} with protocol specifications by quickly exploring various protocol variants.
This process is useful and insightful, as it allows us to observe how modifying the protocol affects its security.
We have incorporated this feature into \easystrands{} through proof automation via Coq tactics that perform case analysis, and eliminate the easy cases, as illustrated in the previous section.
Even though this does not guarantee that proofs can be reused when a specification is modified, in practice, we have observed that it is often the case.
Below, we provide examples supporting this fact.
Moreover, we point out that the proofs for the protocol in \cref{sec:proofprotocol} were mostly reused for the proofs of the NSL protocol, which is entirely different and relies on an asymmetric key cryptosystem (\cref{subsec:NSL}).

\myparagraph{Replacing $A$ with $B$ in the ciphertext}
The role of $A$ in the second protocol message is crucial for the security of the protocol, as it clarifies the direction of the message.
This is attributed to our consideration of the symmetric key $\SK{AB}$ as \emph{bidirectional}, meaning it remains the same whether the protocol is run by $A$ with $B$ or by $B$ with $A$.
Without an identifier in the ciphertext, the protocol would be vulnerable to what is commonly known as a \emph{reflection attack}, which we will discuss in the next section.
Here, we demonstrate that using either $A$ or $B$ in the ciphertext achieves the same result, as both identifiers disambiguate the protocol's direction.
To this aim, we consider a variant where $B$ replaces $A$ in the second message:
\vspace*{-0.2cm}
\begin{align*}
  A \rightarrow B & : A \cdot B \cdot N_a \\
  B \rightarrow A & : \enc{N_a \cdot B}{ \SK{AB}}
\end{align*}
Interestingly, when we make this modification, the security proof of the original protocol remains valid for this variant: we just need to change the ciphertext \lstinline{c} from \lstinline{(⟨ $Na ⋅ $A ⟩_(SK A B))} to \lstinline{(⟨ $Na ⋅ $B ⟩_(SK A B))} and the name of one hypothesis in a single rewrite statement.
This can be attributed to our characterization of the minimal element of the set of nodes using a logical proposition that covers all possible cases, along with the utilization of the \lstinline{simplify_prop} tactic in our proof automation.
This tactic automatically resolves most cases, even if they differ for some terms.
The example can be found in \lstinline{Examples/simple_auth/SimpleAuthWithB.v}.

\myparagraph{A flawed version of the protocol}
If we remove both $A$ and $B$ identifiers from the ciphertext the protocol is subject to a well-known reflection attack.
\begin{align*}
  A \rightarrow B & : A \cdot B \cdot N_a \\
  B \rightarrow A & : \enc{N_a }{ \SK{AB}}
\end{align*}
In this case we can copy-paste the proof of the original protocol to check where and why it fails.
The problem arises in the responder case, which has the goal
\begin{lstlisting}
  SA_responder_strand A B Na [⊖ ($B ⋅ $A) ⋅ $Na; ⊕ c]
\end{lstlisting}
but in the hypotheses, we have
\begin{lstlisting}
  SA_responder_strand B A Na [⊖ ($B ⋅ $A) ⋅ $Na; ⊕ c]
\end{lstlisting}
with \lstinline{A} and \lstinline{B} swapped, indicating a (known) reflection attack where \lstinline{c} is generated by \lstinline{A} itself acting as the responder.
The proof can only be closed when \lstinline{A = B}.
In this particular case, \lstinline{A} is persuaded to communicate with itself, which holds true even if the attacker reflects messages between two distinct sessions.
This example can be found in \lstinline{Examples/simple_auth/SimpleAuthFlawed.v}.

\myparagraph{Relaxing the Term Typing}
A common strategy for aiding automated verification involves constraining term types. In our current example, for example, we assume that \Na belongs to the set \T of atomic terms. A notable advantage of strand spaces lies in the insightful nature of their proofs, allowing the addition of assumptions only when necessary. Consequently, it becomes feasible to establish minimal assumptions for protocol security. This, coupled with our proof automation enabling the reuse of proofs, facilitates experimentation with type relaxation over terms to identify missing assumptions when needed. We conducted such an experiment by relaxing the typing, considering \Na as a general term belonging to \terms, not necessarily atomic,
The first lemma that cannot be proven is the one stating that \lstinline{(SKA A B)} never originates on a honest participant strand. In other words, we cannot prove that honest participants do not leak the symmetric key.
In fact, it might be the case that \Nap, for a given initiator, contains \lstinline{#(SKA A B)} as a subterm.
Therefore, the first restriction we need is:
\begin{lstlisting}
forall U U', ~ #(SK U U') ⊏ Na'
\end{lstlisting}
Intuitively, we impose the requirement that a nonce does not covertly transport the secure key  \lstinline{(SKA A B)} as a subterm. Should this occur, the initial message of the initiator would originate such a key, potentially exposing it to the penetrator.

The second point where the proof for the original protocol fails is in the initiator case of the \lstinline{noninjective_agreement} proposition. At this stage of the proof, we aim to eliminate the possibility that an initiator with parameters \Ap, \Bp, \Nap originates the ciphertext \lstinline{⟨ Na ⋅ A ⟩_(SKA A B)}. Once again, this scenario could arise if this ciphertext is a subterm of \Ap, \Bp, or \Nap.
To address this, we require the following:
\begin{lstlisting}
forall N U U', ~ (⟨ N ⋅ $U ⟩_(SK U U')) ⊏ Na'
\end{lstlisting}
We conclude that the protocol remains secure even when nonces are general terms, as long as they do not covertly transport the secure key and the corresponding ciphertext, the two fundamental ingredients for the security of the protocol.
These conditions are included in the specification of the strands for honest participants.
For example for the initiator (and similarly for the responder):
\begin{lstlisting}
Inductive SA_initiator_strand (A B : T) (Na : 𝔸) :
  Σ -> Prop :=
  | SAS_Init : forall i,
    (forall U U', ~ #(SK U U') ⊏ Na) ->
    (forall N U U', ~ (⟨ N ⋅ $U ⟩_(SK U U')) ⊏ Na) ->
    SA_initiator_strand A B Na
      (i, [ ⊕ $A ⋅ $B ⋅ Na; ⊖ ⟨ Na ⋅ $A ⟩_(SK A B) ]).
\end{lstlisting}

The example can be found in \lstinline{Examples/simple_auth/SimpleAuthUntyped.v}.

\subsection{A New Proof Technique}\label{sec:newproofs}
All proofs are based on the minimality lemma that we described in \cref{sec:background-new} and \cref{ex:proofs}.
However, it is up to the analyst to specify the specific set whose minimal elements exhibit witnesses for certain strands, such as in agreement properties, or whose emptiness proves a particular property, as in secrecy proofs that we will examine next (\cref{subsec:KMP}).

\easystrands{} has allowed us to experiment with various approaches to improve the proof techniques of \cite{FHG98}.
To illustrate this, we consider the dual protocol of \cref{ex:simpleprotocol} in which Alice sends an encrypted nonce to Bob, who decrypts it and sends it back in the clear.
Here, authentication is testified by the unique ability of the responder to decrypt an encrypted random challenge, so there is no ciphertext proving the presence of the responder.
Instead, the fact that the nonce has been decrypted needs to be considered as proof of the presence of Bob. Even here, we need one of the identifiers in the ciphertext to prevent reflection attacks.
The protocol is:
\begin{align*}
  A \rightarrow B & : \enc{N_a \cdot A}{\SK{AB}} \\
  B \rightarrow A & : N_a
\end{align*}
We let \lstinline|c := ⟨ Na ⋅ A ⟩_(SK A B)| and we consider the set of nodes whose term \lstinline|t| satisfies the proposition \lstinline|P := $Na ⊏ t /\ ~ c ⊏ t|.
Intuitively, these nodes contain the nonce \Na but do not contain the ciphertext \lstinline|c|.
Thus, they are $\preceq_C$-preceded by the node where the decryption happens.
Therefore, the minimal element of such a set should identify the responder node that performs the decryption and effectively binds all the responder parameters to the expected values \A, \B, \Na.

This proof technique for encrypted challenges, while effective, has a limitation: to eliminate the penetrator strand that destructs pairs, we need to prove that neither the initiator nor the responder originate pairs \lstinline|g ⋅ h| such that \linebreak\lstinline{c ⊏ h} or \lstinline|c ⊏ g|.
This enables the elimination of the pair destruction case of the penetrator, mainly because the penetrator is the only one that might have generated the problematic pairs containing \lstinline|c| in one element and \Na in the other.
These cases are problematic in general because we could have instances such as \linebreak\lstinline|P g /\ c ⊏ h|, implying \lstinline|~P (g ⋅ h)|. This observation is also mentioned in the proof of the NSL protocol in \cite{FHG98}.

Even though in \easystrands{} we have devised a general lemma to handle these cases uniformly and simply, this property on pairs really depends on the protocol syntax and is unrelated to its security.
Protocols that violate this property cannot be proved secure using this technique.
To overcome this limitation, we have explored a new proof technique, which we call the \emph{protected predicate} technique, and we now illustrate with a variant of the above protocol:
\begin{align*}
  A \rightarrow B & : B \cdot \enc{N_a \cdot A}{\SK{AB}} \\
  B \rightarrow A & : N_a
\end{align*}
This protocol adds \B in the clear in the first message, breaking the requirement that a honest participant strand never originates pairs \lstinline|g ⋅ h| such that \lstinline{c ⊏ h} or \lstinline|c ⊏ g|.
Thus, to prove the security of this protocol we define the following predicate:

\begin{lstlisting}
Fixpoint protected a :=
  match a with
  | $t => t <> Na
  | #_ => True
  | ⟨g ⋅ h⟩_(k) =>
      (k = SK A B /\ g = $Na /\ h = $A) \/
      (protected g /\ protected h)
  | ⟨g⟩_(k) => protected g
  | g⋅h => protected g /\ protected h
  end.
\end{lstlisting}
Intuitively, the condition \lstinline|protected A B Na a| holds if and only if \Na appears in \lstinline|a| in the form \lstinline|⟨ Na ⋅ A ⟩_(SK A B)|, or if it does not appear in \lstinline|a| at all.
Now we consider the set of nodes whose terms do not satisfy this condition and use its minimal element to prove agreement.
In fact,
we can prove that the first node where \Na appears unprotected is the responder node that performs the decryption.

This notion is less demanding than the previous predicate \lstinline|$Na ⊏ t /\ ~c ⊏ t|.
For example, term \lstinline|t = Na ⋅ ⟨ Na ⋅ A ⟩_(SK A B)| does not satisfy \lstinline|$Na ⊏ t /\ ~c ⊏ t| as \lstinline|c ⊏ t|, but satisfies \lstinline|~protected A B Na t| since \Na appears in \lstinline|t| in a form different from \lstinline|c|.
It is easy to see that \lstinline|~protected A B Na g| or \lstinline{~protected A B Na h} imply \lstinline|~protected A B Na (g⋅h)|, which solves the pair destruction case of the penetrator without any extra lemma.
We have used this technique to prove the security of the above protocol, and we have also applied it to the NSL protocol (\cref{subsec:NSL}).

Interestingly, regardless of the proof technique used, it is necessary for this protocol to assume that \lstinline|Na| uniquely originates, even
for
noninjective agreement.
Without this assumption, the attacker could simply guess \lstinline|Na| and impersonate Bob.
In the initial protocol of \cref{sec:proofprotocol}, nonce freshness is only required for injective agreement.
This illustrates the elegance of strand spaces, enabling the distillation of the minimal requirements for security proofs.

The example with the original proof technique can be found in \lstinline{Examples/simple_auth/SimpleAuthDual.v} and the variant using the \lstinline|protected| predicate can be found in \lstinline{Examples/simple_auth/SimpleAuthDualBProtected.v}.

\subsection{Maximal Penetrators and Compositionality}\label{sec:maximal}
We have seen that proofs in \easystrands{} rely on case analysis of the various strands belonging to the penetrator and the honest participants.
The goal is to show that a certain subset of nodes is either empty, as it does not contain a minimal element (e.g., for secrecy), or that it admits a minimal element on a specific honest strand (e.g., for authentication).
For penetrator strands, we typically need to demonstrate that none of them admits a minimal element, thus proving that the penetrator cannot interfere with the desired security property.

While performing our many mechanized proofs, we realized that a more general and efficient way to specify the penetrator would be to take a dual approach.
Instead of listing all possible penetrator strands in the classic Dolev-Yao style, we could define the penetrator in terms of what they cannot do with respect to sensitive cryptographic operations.
In other words, penetrator strands would include all those that do not violate specific cryptographic constraints.
This idea resembles the intriguing approach proposed in \cite{banaSymbolic} to achieve computational soundness results, and, in fact, is commonly used in computational models of cryptography.
Here, we explore this concept in a purely symbolic setting, which, to the best of our knowledge, is novel and unexplored in the literature.

This approach, which we call the \emph{maximal penetrator}, offers several advantages.
First, it allows for proving security without the need to specify a Dolev-Yao attacker, which depends on the specific structure of terms and requires updates whenever new terms, such as cryptographic primitives, are introduced.
Second, it enables the penetrator to be maximized by only specifying what is strictly forbidden in order to achieve the security of a given protocol.
As a result, if the security of two protocols has been proven with respect to their maximal penetrators, they can be composed when they mutually respect each other's maximal penetrator conditions.
Intuitively, given two protocols,
if the behavior of each protocol is fully subsumed by the maximal penetrator of the other, we can safely combine them and derive a security proof for the combined protocol from the individual proofs.
In other words, this approach provides protocol compositionality for free.

We have implemented this technique on the protocol of \cref{ex:simpleprotocol} and its variant presented in \cref{sec:reusing}, where $A$ is replaced by $B$.
We then proved the security of their composition by fully reusing the individual security proofs for each protocol, as explained below.

We begin by defining the concept of maximal penetrator strands.
The key challenge is defining a property that ensures the penetrator does not compromise the cryptographic primitives required for the protocol's security.
Ideally, this property should be minimal in order to maximize the penetrator’s capabilities.
In the simple authentication protocol of \cref{ex:simpleprotocol}, security relies on the ability to encrypt using \lstinline|SK A B|.
Thus, the following definition asserts that encryption by the penetrator should only be allowed if the key is known, i.e., it is readable in cleartext from the network.
\begin{lstlisting}
Definition NoForgeCipher A B n :=
  forall p, originates (⟨ p ⟩_(SK A B)) n ->
    exists n', n' =>+ n /\ term n' = ⊖ #(SK A B).
\end{lstlisting}
The above definition states that for given \A and \B, if a ciphertext \lstinline|⟨ p ⟩_(SK A B)| originates in an output node, there must exist a preceding input node in the same strand where the key \lstinline|SK A B| is read in the clear.

The only other property needed for security is that the penetrator never originates \lstinline|SK A B|.
Therefore, we specify maximal penetrator strands as those whose nodes do not originate \lstinline|SK A B| and satisfy \lstinline|NoForgeCipher A B|:
\begin{lstlisting}
Inductive SA_maximal_penetrator_strand (A B : T) : Σ -> Prop :=
  | SAS_Pen : forall s,
      ( forall n, s = strand n ->
        ~originates #(SK A B) n /\ NoForgeCipher A B n ) ->
        SA_maximal_penetrator_strand A B s.
\end{lstlisting}
Under this maximal penetrator, we were able to prove the same authentication properties that we established using the standard Dolev-Yao penetrator (see \lstinline{Examples/simple_auth/SimpleAuthMaximalEnc.v}).

We then demonstrated several interesting results.
First%
, the Dolev-Yao penetrator is subsumed by the maximal penetrator, confirming that we are not overlooking any significant attacks.
\begin{lstlisting}
Lemma DY_is_SA_maximal_penetrator:
  forall A B s, penetrator_strand (K__P_AB A B) s -> SA_maximal_penetrator_strand A B s.
\end{lstlisting}
This also implies that the results we proved under the Dolev-Yao penetrator can now be derived from those established under the maximal penetrator by simply applying the lemma above.
\ifdefined\COLORDIFF
    \color{cbred}
\else
\fi
Crucially, our maximal penetrator is strictly stronger than the Dolev-Yao penetrator but still allows the protocol to be proved secure.
Consider for example the strand \lstinline{[⊖ ⟨ M ⟩_(SK A B); ⊕ M ]}, where an encrypted message is decrypted without knowledge of the secret key \lstinline{SK A B}.
This strand cannot be constructed by a Dolev-Yao penetrator because such intruders cannot break cryptography.
However, since the strand satisfies \lstinline{NoForgeCipher} and does not originate \lstinline{SK A B}, the maximal penetrator can produce it.
This intuition is formalized in file \lstinline{SimpleAuthMaximalEnc.v} by the lemma \lstinline{SA_maximal_penetrator_not_eq_DY}, whose proof is based on the above example.
\ifdefined\COLORDIFF
    \color{black}
\else
\fi

For compositionality, it is useful to demonstrate that certain honest participants can be mimicked by the maximal penetrator.
In particular, we find that the initiator is always subsumed by the penetrator, as it neither originates ciphertexts nor sensitive keys.
In contrast, for the responder, this holds only when the initiator key \lstinline|SK A' B'| is different from the one the attacker cannot forge, namely \lstinline|SK A B|.
This situation arises when neither \lstinline|A = A' /\ B = B'| nor \lstinline|A = B' /\ B = A'|.
This latter property is also the basis for our compositionality result.
\begin{lstlisting}
Lemma ini_penetrator :
  forall s A B Na A' B',
    SA_initiator_strand A' B' Na s ->
    SA_maximal_penetrator_strand A B s.

Lemma res_penetrator :
  forall s A B Na A' B',
    ~((A = A' /\ B = B') \/ (A = B' /\ B = A')) ->
    SA_responder_strand A' B' Na s ->
    SA_maximal_penetrator_strand A B s.
\end{lstlisting}
We have finally composed the protocol of \cref{ex:simpleprotocol} and its variant presented in \cref{sec:reusing}, where $A$ is replaced by $B$ in the protocol, under the same maximal penetrator.
This is done by simply placing in the same strand space the maximal penetrator strands and the initiator and responder strands of the two protocols, whose identities are required to respectively satisfy two predicates \lstinline|p1| and \lstinline|p2|.
Fixed a maximal penetrator for \lstinline|A| and \lstinline|B|, if we pick
\begin{lstlisting}
Definition p1 (A' B' : T) := True.
Definition p2 (A' B' : T) := ~((A = A' /\ B = B') \/ (A = B' /\ B = A')).
\end{lstlisting}
we allow all participants for protocol 1, as well as all participants with \lstinline|SK A' B'| that is disjoint from \lstinline|SK A B| for protocol 2.
Due to this restriction, protocol 2 can be emulated by the maximal penetrator, thanks to the \lstinline|res_penetrator| lemma.
Consequently, we can prove that the composed strand space is essentially the same as the strand space of protocol 1, allowing us to directly reuse all the results that have already been established for protocol 1.
\begin{lstlisting}
Lemma comp_is_protocol1:
  C_is_SS C (SA_StrandSpace p1 p2 A B) ->
  C_is_SS C (SimpleAuthMaximalEnc.SA_StrandSpace A B).
\end{lstlisting}
The same result also holds for protocol 2 by swapping predicates \lstinline|p1| and \lstinline|p2|.
All details are available in \lstinline{Examples/simple_auth/SimpleAuthMaximalEnc*.v}.

\section{Case Studies}\label{sec:casestudies}
  In addition to the family of simple authentication protocols inspired by the ISO/IEC 9798-2 two-pass authentication protocol (\cref{sec:running}), we have applied \easystrands{} to two nontrivial case studies:
the classic Needham-Schoeder-Lowe protocol and its original flawed version (\cref{subsec:NSL}) \cite{lowe1995attack}, and a recently proposed solution for secure key management policies (\cref{subsec:KMP}) \cite{focardi2021secure}.
Due to space constraints we only briefly describe the highlights and refer the interested reader to the files respectively in \lstinline{Examples/nsl}, \lstinline{Examples/ns_original} and \lstinline{Examples/kmp}.

\subsection{Case Study 1: Needham-Schroeder-Lowe Protocol}\label{subsec:NSL}
The NSL protocol is a standard protocol that has been widely analyzed \cite{lowe1995attack}.
The protocol assumes that $A$ and $B$ know their respective public keys, $\mathit{PK} A$ and $\mathit{PK} B$:
\begin{align*}
    &A \rightarrow B: \langle N_a \cdot A \rangle_{\mathit{PK} B}\\
    &B \rightarrow A: \langle N_a \cdot N_b \cdot B \rangle_{\mathit{PK} A}\\
    &A \rightarrow B: \langle N_b \rangle_{\mathit{PK} B}
\end{align*}
This protocol can be used to mutually authenticate the \emph{initiator} $A$ and the \emph{responder} $B$, while allowing them to share two secret values (the nonces, $N_a$ and $N_b$) that can be used together to generate a shared session key.
Intuitively, the authentication guarantee arises from the fact that only $A$ and $B$ can decrypt the nonces using their private keys and send them back to each other. Meanwhile, cryptography ensures the secrecy of the fresh nonces. This protocol, along with its original flawed version (in which $B$ was absent in the second message), has been used in \cite{FHG98} to illustrate the strand spaces model.

\myparagraph{Results} We have successfully mechanized all the proofs in \cite{FHG98} spotting and fixing two problems, described below. We also applied our new \emph{protected predicate} proof technique from \cref{sec:newproofs} to simplify some proofs.

\myparagraph{First issue}
When proving Lemma 4.4 in \cite{FHG98}, Fabrega et al. consider the set
$T = \{ m \in C \mid m \prec_C n_2 \land g \cdot h \sqsubset \mathit{term} (m) \}$ (for some $n_2$, $g$, and $h$),
and they implicitly assume that $T$ is \emph{sign closed}, i.e., is such that for any pair of nodes $m, m'$ with the same unsigned term, it holds $m \in S$ iff $m' \in S$. However, this is not true when $m \prec_C n_2$ but $m' \not\prec_C n_2$. This means that the authors could not have applied Lemma 2.7 from~\cite{FHG98} that states that minimal elements of sign closed sets are always positive. Fortunately, the conclusion of this lemma still holds when weakening the requirements of being sign closed, and in \easystrands{} we have devised a general lemma to handle these cases uniformly and simply. In particular, we have a lemma similar to 2.7 from~\cite{FHG98} which only requires that for each negative node in a set there exists a preceding positive node that also belongs to the set (see \lstinline|Lemma minimal_is_positive_weak| in \lstinline|Common/Bundles.v|).

\myparagraph{Second issue}
Initiator's nonce secrecy is only sketched in \cite{FHG98}.
Reformulated in Coq, the first part of the initiator's nonce secrecy
\mbox{proposition
from~\cite{FHG98} would
be:}
\begin{lstlisting}
forall m, is_node_of m C -> $Na ⊏ uns_term m ->
    (⟨ $Na ⋅ $A ⟩_(PK B)) ⊏ uns_term m \/
      (⟨ $Na ⋅ $Nb ⋅ $B ⟩_(PK A)) ⊏ uns_term m.
\end{lstlisting}
Intuitively, whenever \Na appears in a node, one of the two above ciphertexts should also appear in the node.
Unfortunately, this proposition fails in (at least) two cases:
$(i)$ consider a node \lstinline{m} that lies on an \emph{initiator} strand with parameters \lstinline{A}, \lstinline{B}, \lstinline{Na}, and \lstinline{Na}.
Then, the third message \lstinline{⟨ $Na ⟩_(PK B)} contradicts the proposition;
$(ii)$ consider \lstinline{m} lying on a \emph{responder} strand with parameters \lstinline{A}, \lstinline{B}, \lstinline{Na}, and \lstinline{Nb' <> Nb}.
Here, the second message \lstinline{$Na ⊏ ⟨ $Na ⋅ $Nb' ⋅ $B ⟩_(PK A)} contradicts the proposition.
To solve this issue and prove the initiator's nonce secrecy, we weakened the theorem just enough by accounting for the missing case \lstinline{(⟨ $Na ⟩_(PK B)) ⊏ uns_term m}, and by letting \lstinline|Nb| free in \lstinline|⟨ $Na ⋅ $Nb ⋅ $B ⟩_(PK A)| 
(full proof in \lstinline{Examples/nsl/NSL_secrecy_initiator_simple.v}).

\subsection{Case Study 2: Key Management Policies}\label{subsec:KMP}
Key management encompasses the practices involved in generating, distributing, storing, and revoking cryptographic keys. To ensure security, keys are commonly stored in tamper-resistant hardware like Hardware Security Modules (HSMs) and accessed through suitable APIs, such as \texttt{PKCS\#11}.
Unfortunately, incorrect key management or overly liberal APIs, which do not allow to provide a policy that precisely determines the intended use of a certain class of keys, may hinder the security of the stored keys ~\cite{anderson00correctness,clulow03pkcs11}.
Among others~\cite{CentenaroFL13,KunPOST15}, Focardi and Luccio \cite{focardi2021secure} proposed security solutions based on {typed key management policies}.
The idea is to dynamically keep track of key types by encrypting a key and its type under a device master key. The policy dictates which key can wrap/unwrap which other key based on the respective types.

The proof in \cite{focardi2021secure} is developed in strand spaces and, due to the overapproximation result, we claim that such general soundness result would be hard if not impossible to achieve using state-of-the-art fully automated tools.
Preliminary tests with Tamarin allowed us to prove the security of specific policies, disregarding the overapproximation part.
Scalability became an issue as the policy size increased,  since the tool had to traverse all policy states for the analysis.

\myparagraph{Results} We fully mechanized the soundness theorem of \cite{focardi2021secure} and uncovered an ambiguous usage of the proof technique in the pen-and-paper development and a redundant case in the original notion of policy closure that we simplified. We improved the precision of the analysis by providing a more accurate closure operation, which allowed us to prove the security of the \emph{secure templates} example \cite{BCFS-ccs10}, previously rejected by the analysis in \cite{focardi2021secure}.

\myparagraph{First issue} The security theorem presented in \cite{focardi2021secure}
is a soundness result.
It establishes that the policy closure overapproximates the key types at runtime and at all bundle nodes.
Focardi and Luccio achieve this by considering the dual set of nodes violating the properties and demonstrating its emptiness through an inductive examination of all possible strands. During our analysis, we found that in the pen-and-paper development, the definition of this set did not encompass all possible cases for subterms. To address this, we employed our novel \emph{protected predicate} proof technique, which centers around the \lstinline|protected| predicate as outlined in \cref{sec:newproofs}. This approach inherently covers all subterms by construction and simplify the treatment of pair terms, especially in penetrator strands.

\myparagraph{Second issue}
While developing the proof mechanization we realized that one of the condition in the policy closure (item 5 in Definition 6 of \cite{focardi2021secure}) dealing with decryption operations was never
used in the proof and could be safely removed (see below for more detail).

\myparagraph{Improving the analysis precision}
While mechanizing  the proof by~\citet{focardi2021secure} we realized that the closure operation could be made more precise, simpler and more intuitive.
In the following we briefly present our improved closure operation and show that it is more precise than the original one by validating a particular policy, proposed in \cite{BCFS-ccs10}, that was rejected by the original analysis.

We need to provide more details about the model presented in \cite{focardi2021secure}.
When a key is created, a type is assigned to it and encrypted along with the key under a secret master key $\mkey$ to enforce the policy at execution time.
For example, key $k_1$ of type $\ckey_1$ is  modeled as $\enc{k_1,\ckey_1}{\mkey}$.
Keys can be used to encrypt and decrypt other keys to securely export them out of the device and possibly import them into another one.
These two operations are usually referred to as \emph{wrap} and \emph{unwrap}.
When a key is unwrapped any type admitted by the policy is assigned to the unwrapped key, making it possible to have multiple types for the same key.
This is modeled by creating another ciphertext with the new assigned type, e.g., $\enc{k_1,\ckey_2}{\mkey}$.

A key management policy is specified as a set of directives $\policyold{\ckey_1}{\lenc}{\ckey_2}$ and $\policyold{\ckey_1}{\ldec}{\ckey_2}$ respectively indicating that keys of type $\ckey_1$ can encrypt keys of type $\ckey_2$, and keys of type $\ckey_1$ can decrypt wrapped keys and assign them type $\ckey_2$.
We also let $D$ denote the type for generic data so $\policyold{\ckey_1}{\lenc}{D}$ and $\policyold{\ckey_1}{\ldec}{D}$ indicate that keys of type $\ckey_1$ can perform standard encryption and decryption operations on messages.
Let $\termDKey$ denote the keys originated in the device,
then the key management API strands have the following form:
\begin{description}[leftmargin=6em,style=nextline]
\item [~~\rm Create:] \strand{+\enc{k,\ckey}{\mkey}} with $k \in \termDKey$ uniquely originating
\item [~~\rm Encrypt:] \strand{-m,~ {-\enc{k,\ckey}{\mkey}},~ {+\enc{m}{k}}} if $\policyold{\ckey}{\lenc}{D}$
\item [~~\rm Decrypt:] \strand{-\enc{m}{k},~ {-\enc{k,\ckey}{\mkey}},~ {+m}} if $\policyold{\ckey}{\ldec}{D}$
\item [~~\rm Wrap:] \strand{-\enc{k_1,\ckey_1}{\mkey},~ {-\enc{k_2,\ckey_2}{\mkey}},~ {+\enc{k_1}{k_2}}} if $\policyold{\ckey_2}{\lenc}{\ckey_1}$
\item [~~\rm Unwrap:] \strand{-\enc{k_1}{k_2},~ {-\enc{k_2,\ckey_2}{\mkey}},~ {+\enc{k_1,\ckey_1}{\mkey}}} if $\policyold{\ckey_2}{\ldec}{\ckey_1}$
\end{description}
Intuitively, Create generates a new device key of type $K$, Encrypt and Decrypt perform standard encrypt and decrypt operations on messages if the policy enables them, Wrap and Unwrap model key management operations in which a key encrypts/decrypts other keys along the policy directives.

A {closure operation} applied to the key management policy yields an overapproximation of the types that a particular key may assume during runtime, and a security theorem establishes the soundness of this overapproximation, ensuring that keys never assuming the insecure \emph{Data} type $D$ are guaranteed to remain undisclosed. The set of types that are \emph{reachable} from an initial type $\ckey$ is noted $\R_\ckey$. To compute this set, a new policy denoted by $\Rightarrow$ is defined, extending $\rightarrow$ to overapproximate all possible key types that can be reached when executing the key management APIs.

The original closure of \cite{focardi2021secure} defines $\Rightarrow$ as the smallest relation such that:
\begin{enumerate}
    \item $\policy{\ckey}{l}{\cdkey}$ implies $\policyC{\ckey}{l}{\cdkey}$;
      \label{item1}
    \item $\ckey \in \R_\ckey$;
      \label{item2}
    \item $\policyC{D}{l}{D}$;
      \label{item2bis}
    \item $\policyC{\ckey}{\lenc}{\cdkey}$ and $\policyC{\ckey}{\ldec}{\czkey}$ implies $\czkey \in \R_\cdkey$;
      \label{item3}
      \item $\policyold{\ckey}{\ldec}{\cdkey}$ and $\ckey \in \R_\czkey$ implies $\policyC{\czkey}{\ldec}{\cdkey}$
     \label{item3bis}
  \item $\policyC{\ckey}{\lenc}{\cdkey}$ and ($\ckey \in \R_\czkey$ or $\czkey \in \R_\ckey$) implies $\policyC{\czkey}{\lenc}{\cdkey}$
       \label{item5}
  \item $\policyC{\cdkey}{\lenc}{\ckey}$ and ($\ckey \in \R_\czkey$ or $\czkey \in \R_\ckey$) implies $\policyC{\cdkey}{\lenc}{\czkey}$
       \label{item6}
  \end{enumerate}
Intuitively,  whatever is allowed by $\rightarrow$ is also allowed by $\Rightarrow$ (item \ref{item1}); a type $\ckey$ is always reachable by itself (item \ref{item2}); $D$ can perform any operation over  $D$, in order to account for penetrator's behaviour (item \ref{item2bis}); if a type $\ckey$ can acquire the capability of wrapping $\cdkey$ and then decrypt it as $\czkey$, then $\czkey$ should belong to the types $\R_\cdkey$ that are reachable from $\cdkey$ (item \ref{item3}). Item \ref{item3bis} propagates decryption capability from $K$ to $J$ if $K$ is reachable from $J$. Similarily, items \ref{item5} and \ref{item6} propagate encryption capabilities bidirectionally.

Developing our mechanized proof we first realized that item \ref{item3bis} was unnecessary, as discussed above, and we removed it.
Moreover,
while this closure can be proved to soundly approximate the propagation of key types and so it is enough for security, the last two items look overly conservative and not very intuitive.
We then devised a more accurate closure which replaces original rules from \ref{item3bis} to \ref{item6} with the following:
\begin{enumerate}
    \setcounter{enumi}{4}
\item $\policyC{\ckey}{\lenc}{\cdkey}$ and $\ckey \in \R_\czkey$ and $\cdkey \in \R_\cwkey$ implies $\policyC{\czkey}{\lenc}{\cwkey}$
     \label{item5b}
\item \vspace*{-.2cm}$\policyC{\ckey}{\ldec}{\cdkey}$ and $\ckey \in \R_\czkey$ implies $\policyC{\czkey}{\ldec}{\cdkey}$
     \label{item6b}
\end{enumerate}
Intuitively, when $K$ and $J$ can be reached by $Z$ and $W$, the encryption capabilities between $K$ and $J$ are inherited by $Z$ and $W$ (item \ref{item5b}).
Similarly, for decryption, the capability to decrypt to a type $J$ is inherited from $K$ by $Z$ if $K$ is reachable from $Z$ (item \ref{item6b}).
These two rules model more accurately the fact that encryption and decryption capabilities are acquired when a certain type $K$ is reached by another type $Z$.
\begin{figure}[t]
    \centering
    \begin{tikzpicture}[node distance=10mm and 20mm, main/.style = {semithick, inner sep=0, draw, circle,minimum width =0.7cm}]
    \node[main] (1) {$\ckey_1$};
    \node[main] (2) [below= of 1]{$\ckey_2$};
    \node[main] (3) [below= of 2]{$\ckey_3$};
    \node[main] (4) [right= of 2]{$D$};
    \path (1) edge [semithick, loop above, ->]  node[midway, above] {\lenc} (1);
    \path (1) edge [semithick, ->]  node[midway, right] {\lenc/\ldec} (2);
    \path (3) edge [semithick, bend right, ->]  node[midway, right] {\lenc/\ldec} (4);
    \draw[->] (1) [semithick, bend right, ->] to[in=225]++ (-2,-2) to[out=315]  node[at start, left] {\lenc} (3);
    \draw[semithick, ->] (2) -- node[midway, above] {\lenc} (4);
    \path (2) edge [semithick, loop left, ->]  node[midway, left] {\ldec} (2);
    \end{tikzpicture}
  \caption{Secure templates of \cite{BCFS-ccs10} as specified in \cite{focardi2021secure}.}
  \label{fig:templates}
\end{figure}
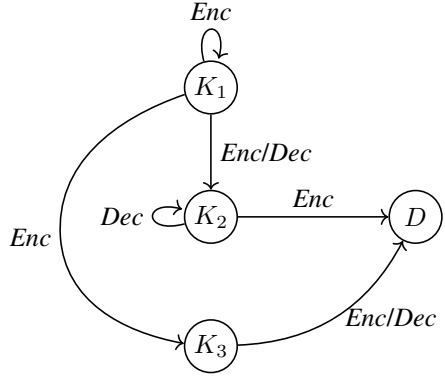

By largely reusing the mechanization of the original proof, we were able to demonstrate that this closure is also sound.
We applied it to all the examples in \cite{focardi2021secure}, reproducing all the results and additionally proving the security of the \emph{secure templates} policy shown in~\cref{fig:templates}.
Intuitively, this policy has a unique type for unwrapped keys ($\ckey_2$) that prevent conflicting roles.
Keys can be generated either as wrap/unwrap keys ($\ckey_1$) or as encrypt/decrypt keys ($\ckey_3$).
When unwrap happens, the imported key assumes type $\ckey_2$ which is only allowed to unwrap and encrypt. The rationale is that unwrap and encrypt operations do not conflict with the initial key roles.

The refined closure provides the following reachable types:
\begin{eqnarray*}
    \R_{\ckey_1} & = & \{ \ckey_1, \ckey_2 \} \\
    \R_{\ckey_2} & = & \{ \ckey_2 \} \\
    \R_{\ckey_3} & = & \{ \ckey_2, \ckey_3 \} \\
    \R_{\ckey_D} & = & \{ \ckey_2, D \}
\end{eqnarray*}
This confirms the intuition that from each type, it is possible to reach only the unwrapped key $\ckey_2$ and nothing else. In turn, this proves the confidentiality of key types $\ckey_1$, $\ckey_2$, and $\ckey_3$, since $D$, the insecure \emph{Data} type, does not appear in their reachable sets. Consequently, values of keys with those initial types will never appear as plaintext. In \cite{focardi2021secure}, instead, it is shown that the original closure computes reachable sets that all contain $D$, making it impossible to draw any conclusions about key confidentiality for this particular policy.
To the best of our knowledge, this is the first proof of security of the \emph{secure templates} policy from \cite{BCFS-ccs10}.

\section{Summary of Analyzed Protocols}
\label{sec:summary}

\renewcommand{\arraystretch}{1.2}
\begin{table*}[tbh]
    \centering
    \begin{tabular}{rrllll}
        &  & \rotatebox[origin=l]{80}{\parbox{3.5cm}{\lstinline|uniquely_originates Na|}} & \rotatebox[origin=l]{80}{\parbox{3.5cm}{\lstinline|SK A B| unknown to the\\ Dolev-Yao penetrator}} & \rotatebox[origin=l]{80}{\parbox{3.5cm}{Maximal penetrator:\\ cannot forge \lstinline|SK A B| and\\ corresponding ciphertexts}} & \rotatebox[origin=l]{80}{\parbox{3.5cm}{\lstinline|B <> Na|}} \\ \hline
       \multirow{3}{*}{\textbf{\begin{tabular}[c]{@{}r@{}}SimpleAuth, SimpleAuthWithB, \\ SimpleAuthUntyped\end{tabular}}} & \lstinline|noninjective_agreement| &  & \checkmark &  \\
        & \lstinline|injectivity| & \checkmark &  &  \\
        & \lstinline|injective_agreement| & \checkmark & \checkmark &  \\ \hline
       \multirow{3}{*}{\textbf{\begin{tabular}[c]{@{}r@{}}SimpleAuthMaximalEnc, \\ SimpleAuthMaximalEncWithB, \\ SimpleAuthMaximalEncComposition\end{tabular}}} & \lstinline|noninjective_agreement| &  &  & \checkmark \\
        & \lstinline|injectivity| & \checkmark &  &  \\
        & \lstinline|injective_agreement| & \checkmark &  & \checkmark \\ \hline
        \multirow{3}{*}{\textbf{SimpleAuthDual}} & \lstinline|noninjective_agreement| & \checkmark & \checkmark &  \\
        & \lstinline|injectivity| & \checkmark &  &  \\
        & \lstinline|injective_agreement| & \checkmark & \checkmark &  \\ \hline
        \multirow{3}{*}{\textbf{\begin{tabular}[c]{@{}r@{}}SimpleAuthDualBProtected\end{tabular}}} & \lstinline|noninjective_agreement| & \checkmark & \checkmark &  \\
        & \lstinline|injectivity| & \checkmark &  &  \\
        & \lstinline|injective_agreement| & \checkmark & \checkmark & & \checkmark  \\ \hline
       \end{tabular}%

    \vspace{1mm}
    \caption{Summary of the requirements of each security property of all the variants of the SimpleAuth from~\cref{sec:running}.}\label{tab:simpleauth}
\end{table*}

\begin{table*}[tbh]
    \centering
    \begin{tabular}{@{}rrllll@{}}
         &  & 
         \rotatebox[origin=l]{80}{\parbox{3.5cm}{\lstinline|uniquely_originates Na|}} & \rotatebox[origin=l]{80}{\parbox{3.5cm}{\lstinline|uniquely_originates Nb| and \lstinline|Na <> Nb|}}
         &
         \rotatebox[origin=l]{80}{\parbox{3.5cm}{\lstinline|inv (PK A)| unknown \\ to the penetrator}} & 
         \rotatebox[origin=l]{80}{\parbox{3.5cm}{\lstinline|inv (PK B)| unknow \\ to the penetrator}}  \\ \midrule
        \multirow{2}{*}{\textbf{Initiator, authentication}} & \lstinline|noninjective_agreement| 
        & \checkmark & & \checkmark & \checkmark    \\
         & \lstinline|injective_agreement_orig| 
         & \checkmark & \checkmark & \checkmark & \checkmark \\ 
         & \lstinline|injectivity| 
         &  \checkmark & &  &   \\
         & \lstinline|injective_agreement| 
         & \checkmark & & \checkmark & \checkmark   \\  \midrule
        \multirow{2}{*}{\textbf{Responder, authentication}} & \lstinline|noninjective_agreement| 
        & & \checkmark & \checkmark &    \\
         & \lstinline|injective_agreement_orig| 
         & \checkmark & \checkmark & \checkmark &   \\
         & \lstinline|injectivity|
          &  & \checkmark  &  & \\
         & \lstinline|injective_agreement| 
         & & \checkmark &  \checkmark & \\  \midrule
        \textbf{Initiator, secrecy} & \lstinline|secrecy_of_Na_neq| & \checkmark &  & \checkmark & \checkmark \\ \midrule
        \textbf{Responder, secrecy} & \lstinline|secrecy_of_Nb_neq| &  & \checkmark  & \checkmark & \checkmark \\ \bottomrule
        \end{tabular}%

    \vspace{1mm}
    \caption{Summary of the requirements of each security property of Needham-Schroeder-Lowe.}\label{tab:nsl}
\end{table*}

We summarize the results we achieved on the various analyzed protocols.

\myparagraph{Simple authentication protocol} \cref{tab:simpleauth} reports the results on all the variants of the simple authentication protocol of \cref{ex:simpleprotocol}.
Variants subject to the same assumptions are grouped in the same row  of the table, with assumptions relating to three security properties: \lstinline|noninjective_agreement|, \lstinline|injectivity|, and \lstinline|injective_agreement|. We first notice that, for all protocol variants, \lstinline|injectivity| never depends on key secrecy, as it is solely related to nonce freshness. In fact, if \lstinline|Na| uniquely originates, then there exists only one unique initiator agreeing on \lstinline|Na|.

In the first row, we consider the protocol variants in which the responder is challenged to encrypt the nonce. They base \lstinline|noninjective_agreement| on the impossibility for the Dolev-Yao penetrator to forge the protocol key \lstinline|SK A B| and require nonce freshness only for \lstinline|injectivity| and \lstinline|injective_agreement|.

In the second row, we consider the proofs of security for \lstinline|SimpleAuth|, \lstinline|SimpleAuthWithB| and their composition, under the maximal penetrator. In this case, \lstinline|noninjective_agreement| holds for all penetrators that cannot forge either \lstinline|SK A B| or the ciphertext but can, for example, decrypt any ciphertexts even without knowing the encryption keys. Interestingly, this confirms that these protocols do not base their security on the secrecy of ciphertexts but only on their integrity. These security results are strictly stronger than the previous ones, which are based on the Dolev-Yao attacker, and for this reason, they enable compositional proofs of security. Nonce freshness is necessary for both \lstinline|injectivity| and \lstinline|injective_agreement|.

For the third and fourth rows, since \lstinline|SimpleAuthDual| and \lstinline|SimpleAuthDualBProtected| rely on decryption, nonce freshness is necessary even for \lstinline|noninjective_agreement| to prevent a trivial attack where the penetrator guesses the nonce and correctly responds to the challenge. As for the previous cases, nonce freshness is necessary for both \lstinline|injectivity| and \lstinline|injective_agreement|. Finally, for \lstinline|SimpleAuthDualBProtected|, the variant in which \lstinline|B| is sent in the clear together with the encrypted challenge in the initiator's message (see \cref{sec:newproofs}), we also need to assume that \lstinline|B <> Na|, since otherwise, the nonce would be leaked by the initiator, breaking authentication. Once again, this highlights the strength of the strand spaces model, which enables reasoning about protocol security and identifying the necessary assumptions for their security.

\myparagraph{Needham-Schroeder-Lowe protocol} \cref{tab:nsl} summarizes the assumptions for the NSL protocol, which align with those in the original paper \cite{FHG98}. Notice that the notion of injective agreement in \cite{FHG98}, noted as \lstinline|injective_agreement_orig| in the table, is in some sense dual to the standard formulation in the literature. For example, for the responder guarantee, it is required that exactly one initiator exists who agrees on \lstinline|A|, \lstinline|B|, \lstinline|Na|, and \lstinline|Nb| (see Proposition 4.8 in \cite{FHG98}). This requires the freshness of \lstinline|Na|, which should not be part of the responder guarantees and, in fact, is not necessary to prevent replay attacks.

Usual formalizations of injective agreement impose the opposite requirement: whenever the responder completes the protocol, at least one initiator must exist, and this initiator's run should match with only one responder run. In other words, there should be exactly one responder for each protocol session (see, e.g., \cite{tamarinmanual}). We have adopted this notion for the simple authentication protocol and have also proved it for NSL, with the corresponding proofs reported in the table under \lstinline|injectivity| and \lstinline|injective_agreement|.

In particular, we observe that \lstinline|injectivity| only requires the freshness of the corresponding nonce, whereas \lstinline|injective_agreement| requires the same assumptions as \lstinline|noninjective_agreement|, unlike \lstinline|injective_agreement_orig|, which is always more demanding. As observed in \cite{FHG98}, we also find that the authentication guarantee for the responder relies solely on the secrecy of the private key of $A$, while the authentication guarantee for the initiator requires the secrecy of both private keys. The same holds for secrecy.

\section{Discussion and Conclusions}

In this paper we have described our efforts in mechanizing the strand spaces framework~\cite{FHG98} in Coq.
To assess the flexibility of the approach and the usability of the library and of the proofs we have analyzed a variety of examples: a basic authentication protocol and some of its variants, the classical Needham-Schroeder-Lowe authentication protocol, and a recent key management API equipped with a key management policy.

Wherever possible, our mechanization remains faithful to the original pen-and-paper development of strand spaces.
At the same time, we put a lot of engineering effort to make the code and the proofs reusable.
For that, we have made the framework modular and parametric in the terms and the penetrator.
Additionally, we have developed a number of strands-specific tactics whose goal is to make the life of the protocol's analyst easier by removing some of the burden of these kinds of proofs.
Indeed, the tactics automate a number of intermediate steps enabling, in some cases, easy proof reuse.
For instance, the proof of the NSL responder's nonce secrecy
 required just one hour of work using the initiator's nonce secrecy.
The mechanization
gives the freedom to experiment with protocols and their properties, while retaining the unique ability of strand spaces-based analyses to give interesting insights on the inner workings of protocols.
With our experiments, we uncovered
and fixed issues, discarded
redundant or unused requirements, and significantly improved previous results on the analysis of key management policies, making it possible to formally prove the security of the \emph{secure templates} policy from \cite{BCFS-ccs10} (\cref{sec:casestudies}).

\cref{tab:simpleauth,tab:nsl} in \cref{sec:summary}  summarize the premises for each security property across the analyzed protocol variants. These premises are essential for our security proofs and offer important insights into the assumptions required to make a security protocol correct. The strand spaces model highlights this aspect, and the use of Coq and the \easystrands{} library further clarifies the minimal and necessary nature of these assumptions, reinforcing the model's ability to accurately capture security requirements.
With the insights from these experiments we also developed a new proof technique which we call \emph{protected predicate} technique that, in certain situations, simplifies the proofs making some previously challenging cases trivial.

Another advantage of having this mechanized platform is that it opens up new and interesting avenues of research.
\ifdefined\COLORDIFF
    \color{cbred}
\else
\fi
For instance, an intriguing enhancement to our framework would be the inclusion of algebraic intruders. We believe they can be implemented using at least two approaches, which we briefly outline below.

Given an equational theory $E$ over a signature $\mathit{FS}$, the first approach requires implementing $E$ as a (terminating and confluent) rewriting system \lstinline{rew_E}, and allow penetrators to use \lstinline{rew_E} to manipulate terms containing symbols of $\mathit{FS}$.
More concretely, we first need to create an instance of \easystrands{} terms with support for function symbols in $\mathit{FS}$, then we can extend the penetrator as:
\begin{lstlisting}
Inductive penetrator_strand : Σ -> Prop := ...
| PT_Eqn : forall (g h : 𝔸) i, replace g h rew_E  -> penetrator_strand (i, [⊖ g; ⊕ h]).
\end{lstlisting}
where \lstinline{replace g h rew_E} holds iff \lstinline{g} can be rewritten as \lstinline{h} under \lstinline{rew_E}.
This approach is inspired by that of Tamarin \cite{MSCB13}.

The second approach aligns  with the method used in DY*~\cite{DY}, where cryptographic primitives are modeled as functions that symbolically represent the actual primitives, e.g., \lstinline{dec (c, k) = (if c = enc (m, k) then m else Error)}.
With these definitions, the equational theory $E$ could be defined using Coq Setoids and used for terms in place of Leibniz equality.
This has the advantage to allow both honest parties and the intruder to transparently use the equational theory.
However, as observed by~\citet{DY}, this approach requires proving (at least) that $E$ is an equivalence relation respected by all functions, predicates, and protocol specifications which can be lengthy and tedious.
\ifdefined\COLORDIFF
    \color{black}
\else
\fi

Despite their age, strand spaces have been a catalyst for extensive research, leading to notable extensions that include authentication tests~\cite{guttman2000authentication}, process algebraic-style choice operators~\cite{YEMMS16},
 compositionality \cite{StrandComposition,StrandIndependence,StrandMixed}, and stateful protocols \cite{J12}.
Many of these advancements are crucial for enhancing the expressiveness and usability of the model.
Our plan is to enhance \easystrands{} by integrating these extensions, thereby enabling scalability to more realistic protocols.
Ultimately, this will help narrow the gap with state-of-the-art tools such as DY* \cite{DY}.
In terms of foundational research, an intriguing avenue involves closely examining the relationship between Paulson's inductive method \cite{Paulson94} and strand spaces. We plan to mechanize Paulson's method in Coq and conduct a comparative analysis to assess the relative merits of these two inductive methods.

Finally, we defined a maximal penetrator as the set of strands that do not violate sensitive cryptographic operations required for protocol security. This method is inspired by the approach in \cite{banaSymbolic} to achieve computational soundness and, to our knowledge, has not been explored in a purely symbolic context before. It allows for proving injective agreement without explicitly defining the Dolev-Yao attacker, which we showed to be \diff{strictly} subsumed by the maximal penetrator. Notably, this approach facilitates the composition of protocols proven secure under their respective maximal penetrators, provided they adhere to each other's constraints. We are currently extending this technique to protocols like NSL, where security relies on decryption capabilities.

\section*{Acknowledgments}
We would like to thank the anonymous reviewers for their comments and suggestions, which greatly helped us in improving this paper.
This work is partially supported by projects ``SEcurity and RIghts In the CyberSpace - SERICS'' (PE00000014 - CUP H73C2200089001), ``Interconnected Nord-Est Innovation Ecoscheme - iNEST'' (ECS00000043 - CUP H43C22000540006), and PRIN/PNRR ``Automatic Modelling and \(\forall\)erification of Dedicated sEcUrity deviceS - AM\(\forall\)DEUS'' (P2022EPPHM - CUP H53D23008130001), all under the National Recovery and Resilience Plan (NRRP) funded by the European Union - NextGenerationEU.

\bibliographystyle{IEEEtranN}

\small{
\bibliography{biblio}
}

\end{document}